\documentclass{aa}
\usepackage{graphicx}
\usepackage{natbib}
\usepackage{txfonts}
\bibpunct{(}{)}{;}{a}{}{,} 

\def\ls{LS~5039}
\def\lsi{LS~I~+61$\degr$303}
\def\psrb{PSR~B1259-63}
\begin{document}

\title{Relativistic Doppler-boosted emission in gamma-ray binaries}
\titlerunning{Relativistic Doppler-boosted emission in gamma-ray binaries}

\author{
Guillaume Dubus 
\and Beno\^it Cerutti 
\and Gilles Henri
}
\authorrunning{Dubus, Cerutti, Henri}
\institute{
Laboratoire d'Astrophysique de Grenoble, UMR 5571 Universit\'e Joseph Fourier Grenoble I / CNRS, BP 53, 38041 Grenoble, France 
}

\date{Draft \today}
\abstract
{Gamma-ray binaries could be compact pulsar wind nebulae formed when a young pulsar orbits a massive star. The pulsar wind is contained by the stellar wind of the O or Be companion, creating a relativistic comet-like structure accompanying the pulsar along its orbit.}
{The X-ray and the very high energy ($>$100 GeV, VHE) gamma-ray emission from the binary LS 5039 are modulated on the orbital period of the system. Maximum and minimum flux occur at the  conjunctions of the orbit,  suggesting that the explanation is linked to the orbital geometry. The VHE modulation has been proposed to be due to the combined effect of Compton scattering and pair production on stellar photons, both of which depend on orbital phase. The X-ray modulation could be  due to relativistic Doppler boosting in the comet tail where both the X-ray and VHE photons would be emitted.}
{Relativistic aberrations change the seed stellar photon flux in the comoving frame so Doppler boosting affects synchrotron and inverse Compton emission differently. The dependence with orbital phase of relativistic Doppler-boosted (isotropic) synchrotron and (anisotropic) inverse Compton emission is calculated, assuming that the flow is oriented radially away from the star (\ls) or tangentially to the orbit (\lsi, \psrb).}
{Doppler boosting  of the synchrotron emission in \ls\ produces a lightcurve whose shape corresponds to the X-ray modulation. The observations imply an outflow velocity of 0.15--0.33$c$ consistent with the expected flow speed at the pulsar wind termination shock. In \lsi, the calculated Doppler boosted emission peaks in phase with the observed VHE and X-ray maximum.}
{Doppler boosting is not negligible in gamma-ray binaries, even for mildly relativistic speeds. The boosted modulation reproduces the X-ray modulation in \ls\ and could also provide an explanation for the puzzling phasing of the VHE peak in \lsi. }
\maketitle

\section{Introduction}
Gamma-ray binaries display non-thermal emission from radio to very high energy gamma rays (VHE, $>$100 GeV). Their spectral luminosities peak at energies greater than a MeV. At present, three such systems are known: \psrb\ \citep{Aharonian:2005br}, \ls\ \citep{Aharonian:2005nj} and \lsi\ \citep{Albert:2006wi}. A fourth system, HESS J0632+057 may also be a gamma-ray binary \citep{Hinton:2008eg}. The systems are composed of a O or Be massive star and a compact object, identified as a young radio pulsar in PSR B1259-63. All gamma-ray binaries could harbour young pulsars \citep{Dubus:2006lc}.

Electrons accelerated in the binary system upscatter UV photons from the companion to gamma-ray energies. The Compton scattered radiation received by the observer is anisotropic because the source of seed photons is the companion star. VHE gamma-rays will also produce $e^+e^-$ pairs as they propagate through the dense radiation field, absorbing part of the primary emission. This is also anisotropic.  Both effects combine to produce an orbital modulation of the gamma-ray flux if the electrons are in a compact enough region. This modulation depends only on the geometry. Orbital modulations of the high-energy (HE, $>$100 MeV) and VHE fluxes have indeed been observed. The modulations unambiguously identify the gamma-ray source with the binary \citep{Aharonian:2006qw,Albert:2006wi,Acciari:2008vf}. 

Synchrotron emission can dominate over inverse Compton scattering at X-ray energies, providing additional information to disentangle geometrical effects from intrinsic variations of the source.  {\em Suzaku} and {\em INTEGRAL} observations of \ls\ have revealed a stable modulation of the X-ray flux  \citep{Takahashi:2008vu, Hoffmann:2008ys}.  Possible interpretations are discussed in \S2. None are satisfying. The key point is that the X-ray flux is maximum and minimum at conjunctions and that this excludes any explanation unrelated to the system's geometry as seen by the observer. 

\begin{figure}
\centering
\resizebox{7cm}{!}{\includegraphics{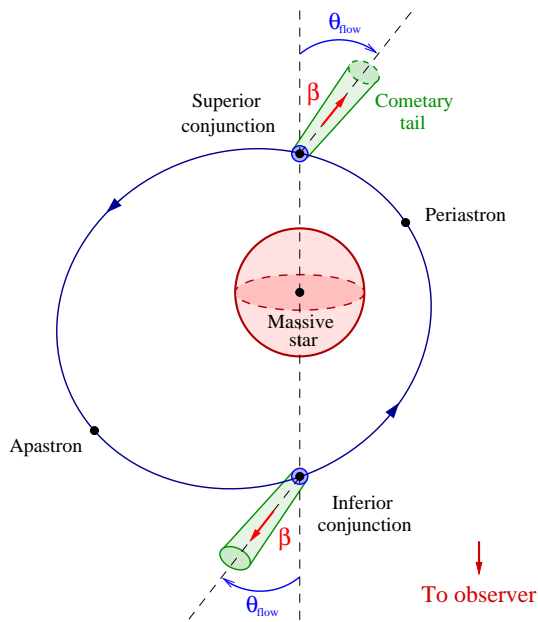}} 
\caption{Geometry of Doppler boosted emission from a collimated shock pulsar wind nebula. The orbit is that of LS 5039 (to scale). The comet tail moves away from the pulsar with a speed $\beta=v/c$ at an angle $\theta_{\rm flow}$. If $\theta_{\rm flow}=0$ then intrinsic emission in the co-moving frame is boosted in the observer frame at inferior conjunction  and deboosted at superior conjunction.}
\label{orb_boost}
\end{figure}

In the pulsar wind scenario, the synchrotron emission is expected to arise in shocked pulsar wind material collimated by the stellar wind. This creates a cometary tail with a mildly relativistic bulk motion (Fig.~\ref{orb_boost}). 
Relativistic Doppler boosting of the emission due to this bulk motion is calculated in \S3 with details given in Appendix A. The orbital motion leads to a modulation of the Doppler boost, as previously proposed in the context of black widow pulsars \citep{1993ApJ...403..249A,2007A&A...463L...5H}. The calculated synchrotron modulation is similar to that seen in X-rays in \ls. Although this is not formally confirmed due to their long orbital periods, \lsi\ and \psrb\ also appear to have modulated X-ray emission \citep{Chernyakova:2006cu,2009MNRAS.397.2123C,2009ApJ...700.1034A,2009ApJ...706L..27A}.  The application to these gamma-ray binaries is discussed in \S4. 

\section{The X-ray modulation in \ls}	

\subsection{X-ray observations}
LS\ 5039 has shown steady, hard X-ray emission since its discovery  \citep{Motch:1997qk,Ribo:1999qo,Reig:2003eq,Martocchia:2005kq,Bosch-Ramon:2005zc,Bosch-Ramon:2007fq}. {\em RXTE} observations hinted at orbital variability \citep{Bosch-Ramon:2005zc} but confirmation had to wait the {\em Suzaku} and {\em INTEGRAL} observations \citep{Takahashi:2008vu, Hoffmann:2008ys}.  The average spectrum seen by {\em Suzaku} from 0.6~keV to 70 keV is an absorbed power-law with spectral index $\alpha=0.51\pm0.02$ ($F_\nu\sim\nu^{-\alpha}$) and $N_{\rm H}=7.7\pm0.2\times 10^{21}$ cm$^{-2}$ and $F_{\rm 1-10\ keV}= 8\times 10^{-12}$ erg cm$^{-2}$ s$^{-1}$, consistent with previous observations. There is no evidence for a cutoff up to 70 keV.

Variability in {\em Suzaku} is dominated by a well-resolved modulation followed over an orbit and a half. The X-ray flux varies by a factor 2 with a minimum at $\phi\approx 0.1$, slightly after superior conjunction ($\phi_{\rm sup}=0.05$ based on \citealt{2009ApJ...698..514A}) and a maximum at inferior conjunction ($\phi_{\rm inf}=0.67$). The 1--10~keV photon index is also modulated, varying between 1.61$\pm0.04$ at minimum flux and 1.46$\pm 0.03$ at maximum flux.  The comparison with {\em Chandra} and {\em XMM} measurements suggests the modulation is stable on timescales of years \citep{2009ApJ...697L...1K}.  The column density is constant with orbital phase, as if there were only absorption from the ISM. The lack of significant wind absorption suggests that the X-ray source is located far from the system or that the wind is highly ionised and/or has a mass-loss rate $\la 10^{-7} M_\odot {\rm yr^{-1}}$ \citep{Bosch-Ramon:2007fq}. Here, we assume that the X-ray source is situated within the orbital system.

\subsection{Inverse Compton X-ray emission?}
The phases of X-ray and VHE maximum (minimum) are identical. If both are due to inverse Compton scattering off stellar photons then maximum emissivity is at superior conjunction. Subsequent in-system absorption due to pair production moves the observed VHE maximum flux to the inferior conjunction.  X-ray photons are too weak for pair production but could be absorbed in the stellar wind with a similar result. This can be ruled out since the modulation is seen in hard X-rays above 10 keV and $N_{\rm H}$ is constant with orbital phase. Thomson scattering of the hard X-rays is unlikely as it would require a column density of scattering electrons $\approx10^{24}$ cm$^{-2}$ (e.g. a Wolf-Rayet wind as in Cyg X-3 rather than an O star wind), two orders-of-magnitude above the observed absorbing column density and plausible stellar wind column densities. 

\begin{figure}
\centering
\resizebox{\hsize}{!}{\includegraphics{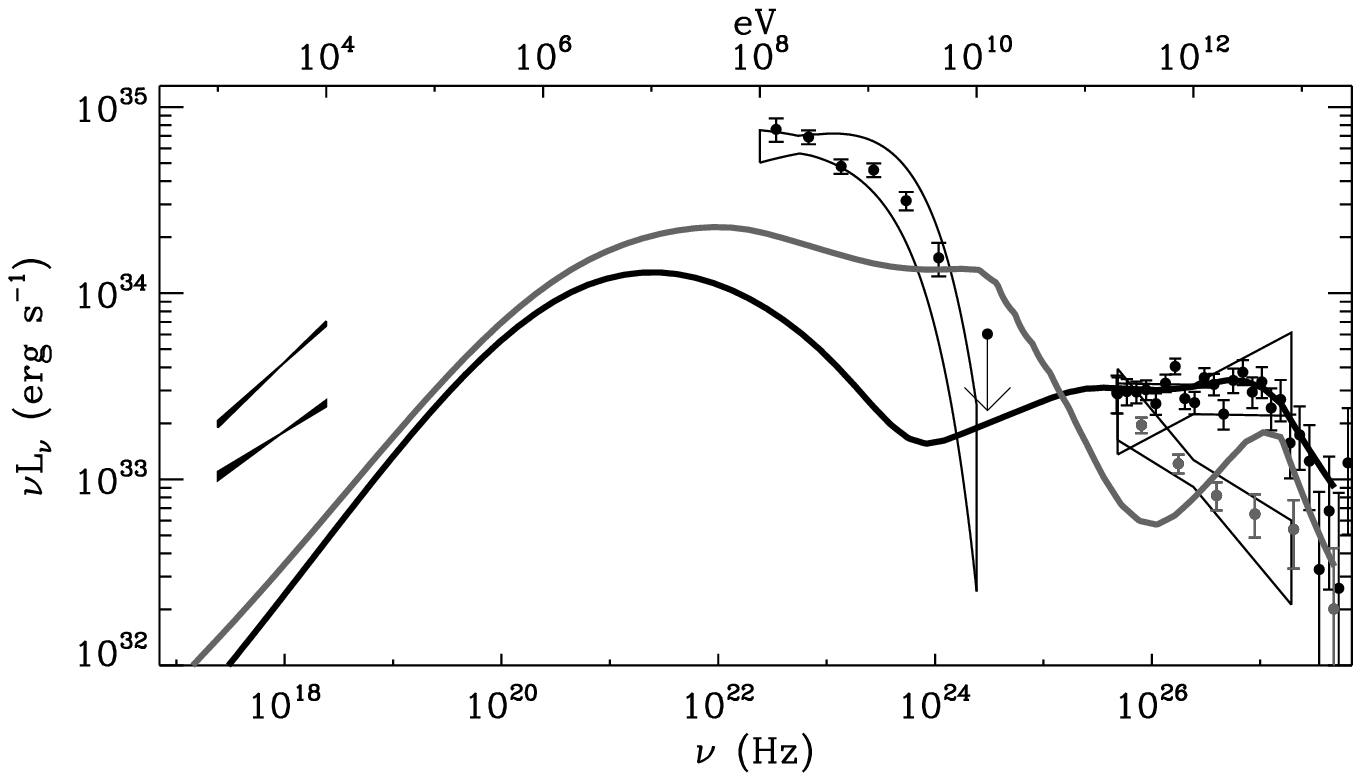}} 
\resizebox{\hsize}{!}{\includegraphics{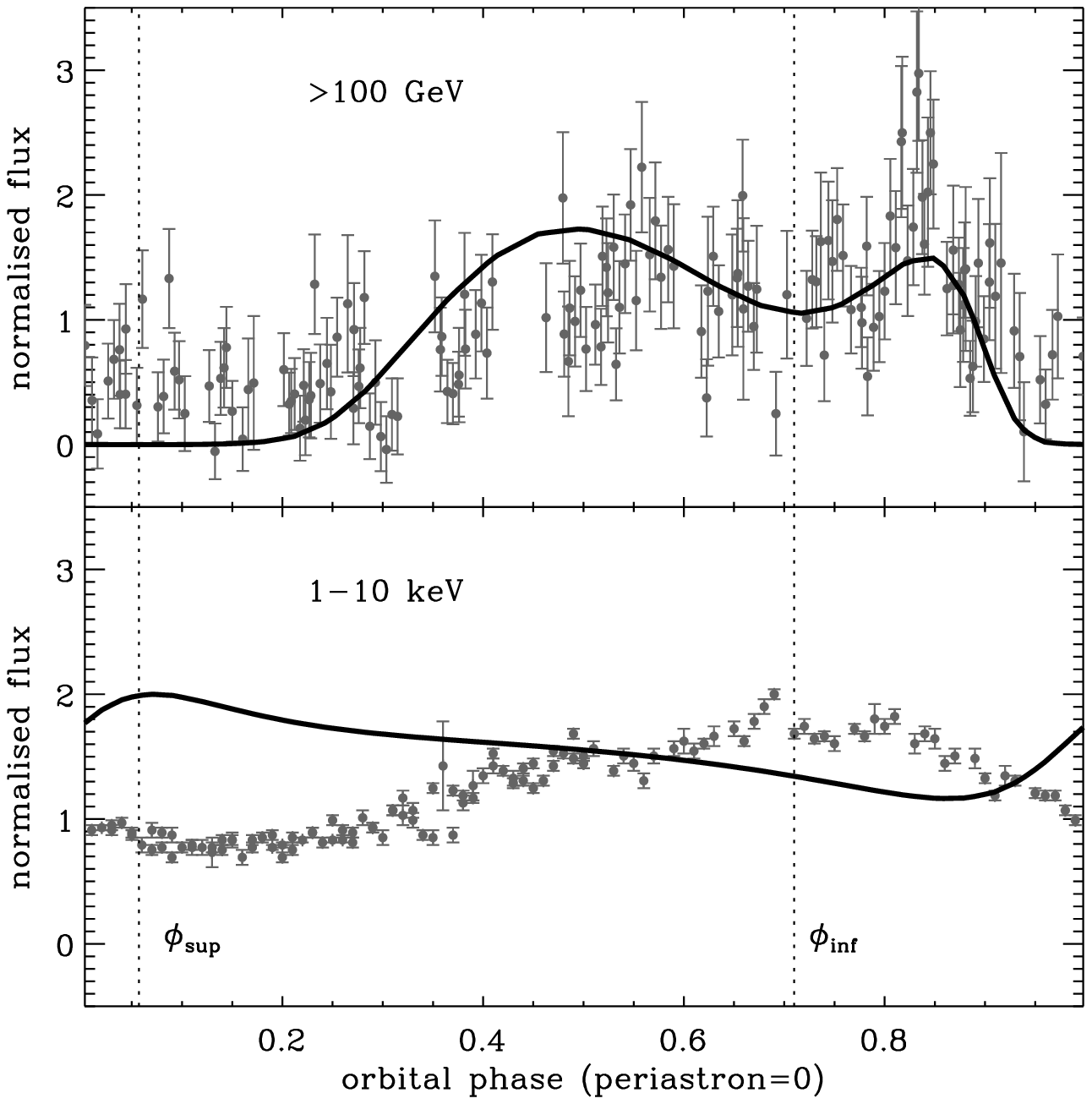}} 
\caption{Comparison of the model for LS 5039 described in \S2.3 with observations. Top panel: spectral energy distribution showing the {\em Suzaku} 1-10 keV maximum and minimum  spectra \citep{Takahashi:2008vu}, the 100 MeV - 10 GeV  average {\em Fermi} spectrum \citep{2009ApJ...706L..56A} and the VHE spectra averaged over phases INFC (dark points) and SUPC (grey points) as defined in \citet{Aharonian:2006qw}. The model spectra averaged over INFC and SUPC are shown as dark and grey curves. Middle and bottom panels: expected VHE gamma-ray and X-ray orbital modulation compared to the HESS and {\em Suzaku} observations.}
\label{sedlc}
\end{figure}

\subsection{Synchrotron X-ray emission?}
Alternatively, the X-ray emission is synchrotron radiation from the same electrons that emit HE and VHE $\gamma$-rays.  In \citet{Dubus:2007oq}, we proposed that several  features of the VHE observations could be explained by assuming continuous injection of a $E^{-2}$ power-law of electrons at the location of the compact object in a zone with a homogeneous magnetic field $B$ of order 1 G \citep{Dubus:2007oq}.  
The synchrotron X-ray spectrum expected in this model\footnote{Here, the injected number of fresh particles is kept constant along the orbit whereas the energy density of cooled particles had been kept constant in \citet{Dubus:2007oq}. With the energy density constant, the particle distribution varied very little with orbital phase, which highlighted the impact of anisotropic scattering. However, a constant injection in number of particles is probably more realistic (at least for a pulsar wind). It has no noticeable influence on the spectra but it slightly changes the VHE lightcurve from that shown in  \citet{Dubus:2007oq}. The VHE lightcurve remains compatible with the HESS results.} is shown in Fig.~\ref{sedlc}. It is hard with $\alpha \approx 0.5$. The electrons producing this X-ray synchrotron emission have energies  between 10 GeV and 1 TeV, for which the dominant cooling mechanism is inverse Compton scattering in the Klein-Nishina regime. This keeps the steady-state distribution close to the $E^{-2}$ power law (Fig. 3 in \citealt{Dubus:2007oq}). Synchrotron cooling takes over at higher energies, causing a break to $\alpha \approx 1$. In fact, the spectral index seen by {\em INTEGRAL} up to 200 keV is softer ($\alpha=1\pm0.2$) than the average index measured by {\em Suzaku}  up to 70 keV  ($\alpha=0.51\pm0.02$). 

Whereas it is promising to have the hard X-ray spectral shape correctly reproduced, the level of X-ray emission is too low and, more importantly, the orbital X-ray lightcurve from the model is inconsistent with the observed modulation. The expected 1-10~keV lightcurve shows only a very modest change with a peak at periastron (Fig.~\ref{sedlc}).  The reason is that the variations in particle and magnetic energy densities (a factor 4) compensate to keep the synchrotron emission almost constant.

\subsection{Variations in parameters?}
A better fit is possible by treating $B$ or particle injection as free functions of orbital phase or by taking adiabatic losses into account. \citet{Takahashi:2008vu} argued that the X-ray spectrum necessarily implies dominant adiabatic cooling of an $E^{-2}$ electron distribution (this is sufficient but not necessary: as discussed above, Klein-Nishina cooling also keeps the distribution hard). The X-ray and VHE observations were then be fitted by adjusting the adiabatic timescale $t_{\rm ad}$ with orbital phase. The derived variation in $t_{\rm ad}$ mirrors the X-ray lightcurve with $t_{\rm ad}$ reaching a maximum at $\phi_{\rm inf}$. There is no obvious reason why $t_{\rm ad}$ should peak at this phase. \citet{Takahashi:2008vu} expect the variation in  $t_{\rm ad}$ to reflect variations in the size of the emitting zone, itself modulated by the external pressure of the wind. The relevant phases are those of apastron (low pressure) and periastron (high pressure), but not inferior conjunction which is an observer-dependent phase unrelated to wind pressure. In LS 5039, $\phi_{\rm inf}$ is significantly different from the phases of periastron and apastron passage. Hence, it would require a coincidence for any intrinsic change in the source ($B$, number of particles, $t_{\rm ad}$, size, etc) to result in a peak at this conjunction.\\

The link between the extrema of the X-ray lightcurve and conjunctions calls for a {\em geometrical} explanation related to how the observer views the X-ray source. Doppler boosting (see Fig.~\ref{orb_boost})  is a possible solution to this puzzle.
 
\section{Relativistic Doppler boosting}
In the interacting winds scenario, the X-ray emission is expected to occur beyond the shock where the ram pressures balance \citep{1977A&A....55..155B,Maraschi:1981nj,Tavani:1994qu,Dubus:2006lc}. Particles in the shocked pulsar wind are randomized and accelerated. MHD jump conditions for a perpendicular shock and a low magnetisation pulsar wind give a post-shock flow speed of $c$/3 \citep{Kennel:1984pd}. If the ratio of wind momenta $\eta=(\dot{E}_{\rm p}/c)/(\dot{M}_\star v_\star)$ is small then the shocked pulsar wind is confined by the stellar wind. The shocked wind flows away from the companion star forming a comet-like tail of emission. Relativistic hydrodynamical calculations show the flow is conical with an opening angle set by $\eta$ and can reach highly relativistic speeds \citep{2008MNRAS.387...63B}. High energy electrons emit VHE gamma-rays and synchrotron X-rays close to the pulsar and lose energy as they flow out, emitting in the radio band far from the system \citep{Dubus:2006lc}. Here, the relativistic electrons radiating X-rays (by synchrotron) and VHE $\gamma$-rays (by inverse Compton) are assumed to be localized at the compact object location. The calculation of the relativistic Doppler boosting in the flow is general and can also be applied e.g. to the case of  a relativistic jet in a binary \citep{2010arXiv1002.3888D}.

\subsection{Synchrotron}
\label{sync}
Even if the flow is only mildly relativistic, Doppler boosting can introduce a geometry-dependent modulation of emission that is isotropic in the comoving frame (Fig.~\ref{orb_boost}). This will be the case for synchrotron emission. The relativistic boost is given by 
\begin{equation}
{\cal D}_{\rm obs}=\frac{1}{\Gamma(1- \beta \mathbf{e_{\rm obs}}.\mathbf{e_{\rm flow}})}
\label{dobs}
\end{equation}
where  $\mathbf{e_{\rm flow}}$ is the unit vector along the direction of the flow and $\mathbf{e_{\rm obs}}$ is the unit vector from the emission site, assumed to be the compact object location, to the observer.  The flow will be assumed to be in the orbital plane where it   makes an angle $\theta_{\rm flow}$ to the star - compact object direction.

The outgoing energy will be modified by $\epsilon={\cal D}_{\rm obs}\epsilon^\prime$ and the outgoing flux will be $F_\nu(\epsilon)={\cal D}^3_{\rm obs}  F^\prime_\nu(\epsilon^\prime)$, with primed quantities referring to the comoving frame. 
In the case of a constant synchrotron power-law spectrum in the comoving frame with index $\alpha$ then 
\begin{equation}
F_{\rm syn}\propto {\cal D}_{\rm obs}^{3+\alpha} 
\label{fsyn}
\end{equation}
The ratio of maximum to minimum flux is \citep[see also][]{1987ApJ...319..416P}
\begin{equation}
\frac{F_{\rm max}}{F_{\rm min}}=\left(\frac{1+\beta\sin i}{1-\beta\sin i}\right)^{3+\alpha}\approx 8 
\label{amplitude}
\end{equation}
for $\beta$=1/3, $i$=60\degr, $\alpha$=0.5. Relativistic boosting can significantly change the theoretical X-ray lightcurve discussed in \S2. In the case of a purely radial flow ($\theta_{\rm flow}$=0$\degr$), maximum (minimum) boost occurs at the inferior (superior) conjunction ($\psi_{\rm obs}=\pi/2-i$ or $\pi/2+i$) where the flow is directed towards (away from) the observer.

\subsection{Inverse Compton}
Inverse Compton emission will also be modified by relativistic aberration. The spectrum of the target photons seen in a given solid angle in the comoving flow frame will be changed according to a different relativistic transform. If the star is assumed to be point-like, the relativistic boost involved is
\begin{equation}
{\cal D}_{\star}=\frac{1}{\Gamma(1- \beta \mathbf{e_{\star }}.\mathbf{e_{\rm flow}})}
\end{equation}
 The total energy density from the star  in the flow frame is
 \begin{equation}
 u_\star={\cal D}_{\star}^{-2} \pi  \left(\frac{R_\star}{d}\right)^2 \frac{a_{\rm SB} T_\star^4}{4\pi}
 \end{equation}
with the Stefan-Boltzmann constant $a_{\rm SB}=7.56\ 10^{-15}$ erg\ cm$^{-3}$ K$^{-4}$. The angle $\psi$ under which scattering occurs will also be changed. This angle ($\cos\psi^\prime=\mathbf{e^\prime_{\star}}.\mathbf{e^\prime_{\rm obs}}$) is given in Appendix A. The inverse Compton spectrum is  then calculated in the comoving frame as in \citet{Dubus:2007oq}. The resulting spectrum is then transformed back to the observer frame as in \S\ref{sync}.
 
 Because of this double transform, and because of the intrinsic orbital phase dependence of scattering on stellar photons,  the Doppler-boosted inverse Compton flux variability can be quite different from the Doppler-boosted synchrotron variability. In the case of Thomson scattering off a power-law of electrons $dN\propto\gamma^{-p}d\gamma$ (see Appendix A)
\begin{equation}
F_{\rm ic}\propto {\cal D}_{\rm obs}^{3+p}(1-\mathbf{e_\star}.\mathbf{e_{\rm obs}})^{\frac{p+1}{2}}  d^{-2}
\label{fic}
\end{equation}
Note that $F_{\rm ic}$ takes into account the decrease in target photon density with distance $d$ to the star since the orbits are not circular. Test calculations show this approximation captures the main features of the full calculation at high energies, including in the Klein-Nishina regime (see also \citealt{2001ApJ...561..111G}). It will be used to discuss the behaviour of the inverse Compton emission.  

\section{Discussion}
\begin{figure}
\centering
\resizebox{\hsize}{!}{\includegraphics{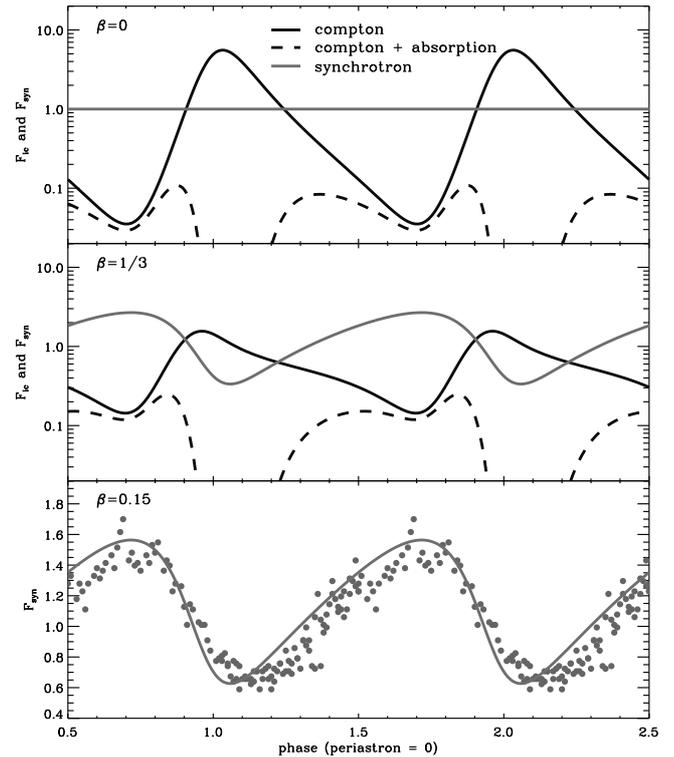}} 
\caption{Doppler-boosted synchrotron ($F_{\rm syn}$, Eq.~\ref{fsyn}) and inverse Compton ($F_{\rm ic}$, Eq~\ref{fic}) intensity variations  for \ls\ assuming $\beta$=0 (top), $\beta=1/3$ (middle), $\beta$=0.15 (bottom). In all panels $\alpha$ is $0.5$ (equivalent to $p$=2) as given by the X-ray spectrum. The flow direction is radial ($\theta_{\rm flow}=0\degr$). Dashed lines show $F_{\rm ic}$ after attenuation due to pair production at 1 TeV. The bottom panel shows a comparison of $F_{\rm syn}$ with the {\em Suzaku} X-ray measurements of \citet{Takahashi:2008vu}. The X-ray data is multiplied by a constant renormalization factor and $\beta$=0.15 to match the X-ray amplitude. }
\label{fig_boost}
\end{figure}

\begin{figure}
\centering
\resizebox{\hsize}{!}{\includegraphics{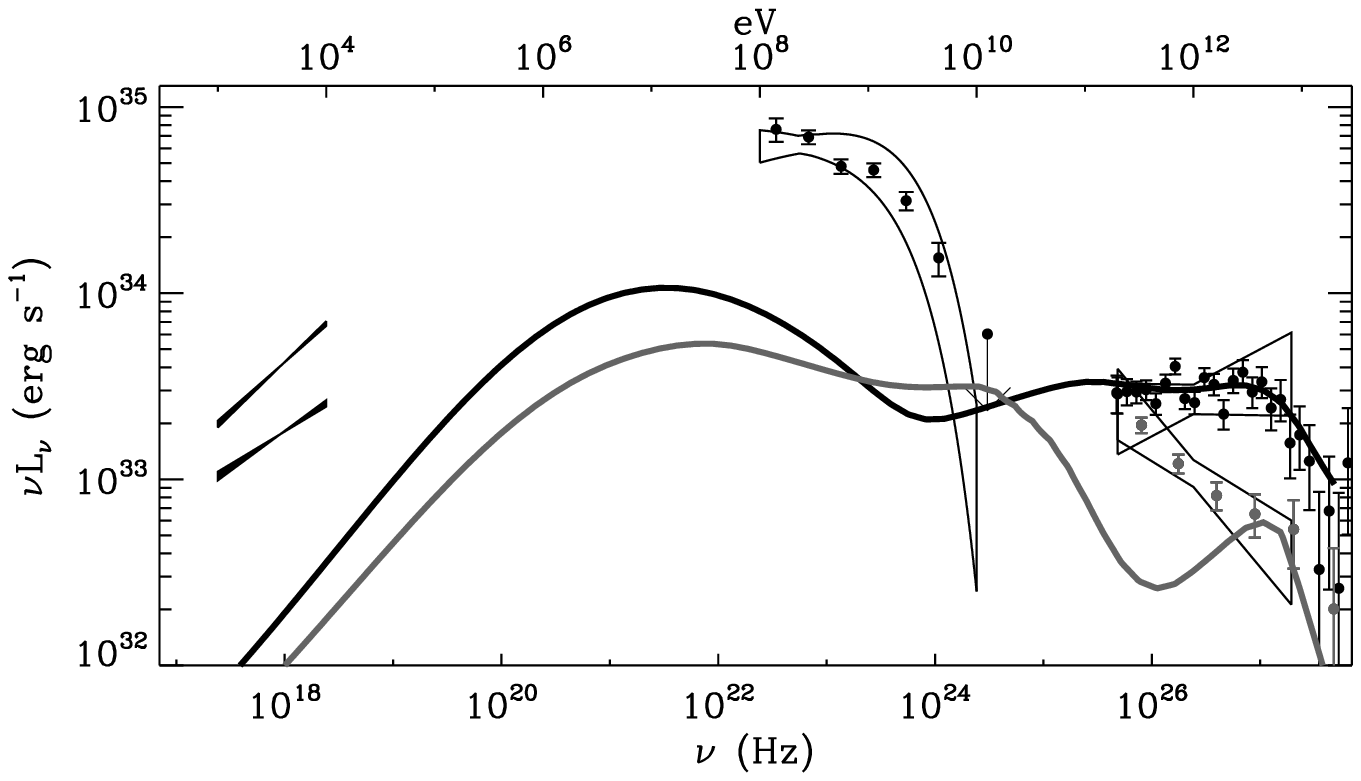}} 
\resizebox{\hsize}{!}{\includegraphics{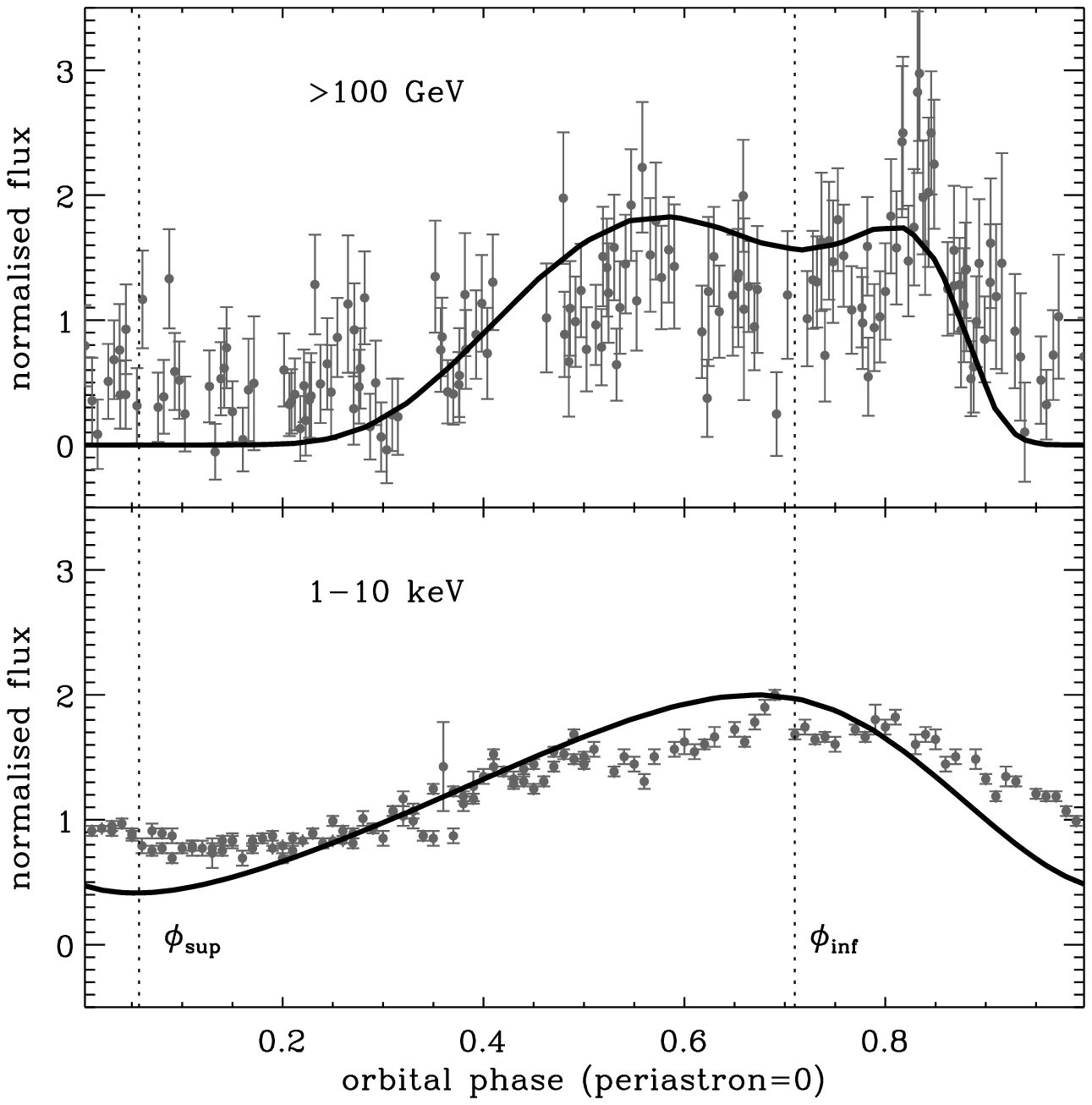}} 
\caption{Same as Fig.~1 but with the corrections due to relativistic motion taken into account. The flow is assumed to originate at the compact object location with $\beta=1/3$ and to point radially outwards from the star.}
\label{sedlc_boost}
\end{figure}

The Doppler-boosted synchrotron ($F_{\rm syn}$, Eq.~\ref{fsyn}) and inverse Compton ($F_{\rm ic}$, Eq.~\ref{fic}) intensity variations were calculated for the three gamma-ray binaries and are discussed here. Full calculations were also carried out for \ls\ and \lsi. The orbital parameters are taken from \citet{Manchester:1995ck} for \psrb\ and from \citet{2009ApJ...698..514A} for \ls\ and \lsi. The inclination $i$ is assumed to be $i$=30$\degr$ for \psrb, and 60$\degr$ for both \ls\ and \lsi\ \citep{Dubus:2006lc}.

\subsection{LS 5039}
\ls\ has a  stellar wind velocity ($v_w\approx 2500$ km s$^{-1}$) significantly greater than the compact object orbital velocity ($v_{\rm orb}\leq 400$ km s$^{-1}$) so that the cometary flow is assumed to be purely radial ($\theta_{\rm flow}=0$). Doppler boosting leads to peaks and troughs for the synchrotron emission $F_{\rm syn}$ at conjunctions as outlined in \S3 (Fig.~\ref{fig_boost}).  The amplitude of the inverse Compton flux ($F_{\rm ic}$) is reduced as the increased scattering rate at superior conjunction is compensated by a deboost of ${\cal D}_{\rm obs}$ (and vice-versa at inferior conjunction). The shape of the modulation does not change much. The bottom panel shows that $F_{\rm syn}$ follows well the {\em Suzaku} data when $\beta$ is adjusted to 0.15 in order to match the X-ray modulation amplitude. The spectral index is fixed to the value observed by {\em Suzaku}, $\alpha=0.5$ (equivalent to $p$=2 for the electron distribution). However, this assumes the intrinsic synchrotron  emission is constant with orbital phase, unlike what happens in the model discussed in \S2 and shown in Fig.~\ref{sedlc}.

The precise relativistic corrections were applied to the model discussed in \S2 (Fig.~\ref{sedlc}), assuming $\beta=1/3$. No other changes were made. The average level of X-ray emission is not changed much. However, the relativistic corrections move the peak X-ray flux to superior conjunction and increase the amplitude of the variations, bringing the model X-ray lightcurve very close to the observations (shown in the bottom panel of Fig.~\ref{sedlc}). The spectral shape is slightly harder than the observed one by about 0.15 in the index $\alpha$. The orbital modulation of $\alpha$ follows the X-ray lightcurve with a hardening of $\alpha$ from 0.42 (superior conjunction) to 0.30 (inferior conjunction), which is similar in amplitude to the hardening observed by {\em Suzaku} (\S2.1). However, the average level of X-ray emission is systematically too low compared to the observations. Increasing the magnetic field by a factor 3 would be sufficient to raise the level of X-ray flux but this would also modify the VHE spectrum, bringing the break at a few TeV to energies that are too low. The model assumes all the emission arises within a single zone and this could explain this shortcoming. The X-ray (and GeV) emission come from electrons that have significantly cooled since their injection and, thus, this emission would be more likely to be affected by a more detailed model where particle cooling is followed along the flow, as done in  \citet{Dubus:2006lc} based on the \citet{Kennel:1984pd} model for pulsar wind nebula. Numerical simulations are needed to provide detailed constraints on the geometry and physical conditions in the post-shock flow.

As expected, the VHE gamma-ray lightcurve is not affected much by the corrections because most of the escaping VHE gamma-rays are emitted close to inferior conjunction (as a result of the $\gamma\gamma$ opacity). The modified VHE spectrum for SUPC phases is actually better than  the original model that overestimated the VHE flux at a few TeV. Pair cascading can fill in the flux between 30 GeV and a few TeV at this phase (Cerutti et al., submitted). At HE gamma-ray energies, in the {\em Fermi} range, the average flux level is reduced significantly because most of the HE gamma rays arise at superior conjunction where the flow deboosts the emission. {\em Fermi} observations of \ls\ and \lsi\ show that the HE gamma-ray emission cuts off exponentially at a few GeV, suggesting the emission in the {\em Fermi} range (100 MeV - 10 GeV) is a distinct component from the shocked flow  \citep{2009ApJ...701L.123A,2009ApJ...706L..56A}. This could be due to pulsar magnetospheric emission, in which case the relativistic corrections and model discussed here will not apply to the GeV component.

\subsection{\lsi}
\begin{figure}
\centering
\resizebox{\hsize}{!}{\includegraphics{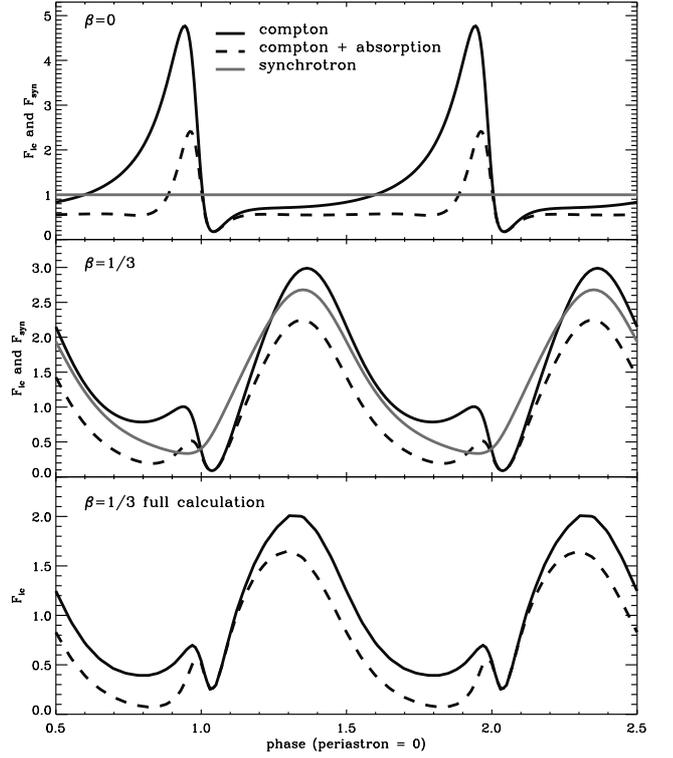}} 
\caption{Doppler-boosted synchrotron ($F_{\rm syn}$, Eq.~\ref{fsyn}) and inverse Compton ($F_{\rm ic}$, Eq.~\ref{fic}) intensity variations  for \lsi\ assuming $\beta$=0 (top) and 1/3 (middle). Here, the direction of the flow is assumed to be tangent to the orbit ($\theta_{\rm flow}\neq0$). The bottom panel shows the inverse Compton emission above 100 GeV using the full calculation instead of Eq~\ref{fic}. Dashed lines show the inverse Compton emission after absorption due to pair production. In all panels a $p$=2 power-law of electrons,  corresponding to $\alpha$=0.5 for synchrotron, is assumed. An offset of 0.275 should be added to the above phases to compare with the  radio-based ephemeris of \lsi.}
\label{fig_boost_b}
\end{figure}
The impact of the relativistic Doppler corrections in \lsi\  (and \psrb) is more difficult to assess because the orientation of the cometary flow is uncertain. The wind of the Be stellar companion is thought to be composed of a fast, tenuous polar wind and, more prominently, a slow, dense equatorial wind. These equatorial winds are effectively Keplerian discs with a small outflow velocity (compared to their angular velocity). If the compact object moves through this disc, then (neglecting corrections due to the orbital eccentricity) it is essentially moving through a static medium in the corotating frame, suggesting the outcome is more likely to be cometary flow trailing the orbit rather than directed radially away from the companion star. This will have to be confirmed by numerical simulations of the interaction.

VHE observations by the MAGIC and VERITAS collaborations consistently find that the peak VHE emission occurs at phases 0.6-0.7 using the historical radio ephemeris \citep{Acciari:2008vf,2009ApJ...693..303A}. The best estimation of the periastron  passage phase in this ephemeris is 0.275 \citep{2009ApJ...698..514A}, hence there is an offset of 0.275 between the radio ephemeris used by observers and the one used here. As outlined in \S2, the phases of periastron/apastron passage or the conjunctions are the natural phases where the physical conditions or the configuration of the system would be expected to produce minima or maxima in the lightcurves. The peak VHE flux occurs 2 to 5 days before apastron and is clearly not associated with any of those phases, making it difficult to interpret only with anisotropic inverse Compton emission and pair production. 

Superior conjunction in \lsi\ occurs slightly before periastron passage, and inferior conjunction slightly after. The inverse Compton peak and trough match exactly with the conjunctions when there is no correction (top panel, Fig.~\ref{fig_boost_b}). Doppler corrections have a strong impact on the inverse Compton lightcurve. Figure~\ref{fig_boost_b} shows the correction factors for  \lsi\  if the flow velocity vector is taken to be exactly tangent to the orbit. The maximum boost is around phases 0.3-0.4 and the emission is deboosted around periastron passage. The effect is strong enough to push the maximum of $F_{\rm ic}$ and $F_{\rm syn}$ at phases 0.57-0.67, using the radio ephemeris, as observed. The correlated behaviour is also consistent with the X-ray and VHE observations reported in \citet{2009ApJ...706L..27A}. These conclusions also hold when doing a full calculation (bottom panel, Fig.~\ref{fig_boost_b}) to properly take into account the Klein-Nishina cross-section. The calculation assumed a constant power-law distribution of electrons with $p$=2. The VHE spectrum is $F_\nu\sim \nu^{-2}$ because of Klein-Nishina effects and the X-ray spectrum is $F_\nu\propto \nu^{-0.5}$, both of which agree with observations.

\subsection{\psrb}

\begin{figure}
\centering
\resizebox{\hsize}{!}{\includegraphics{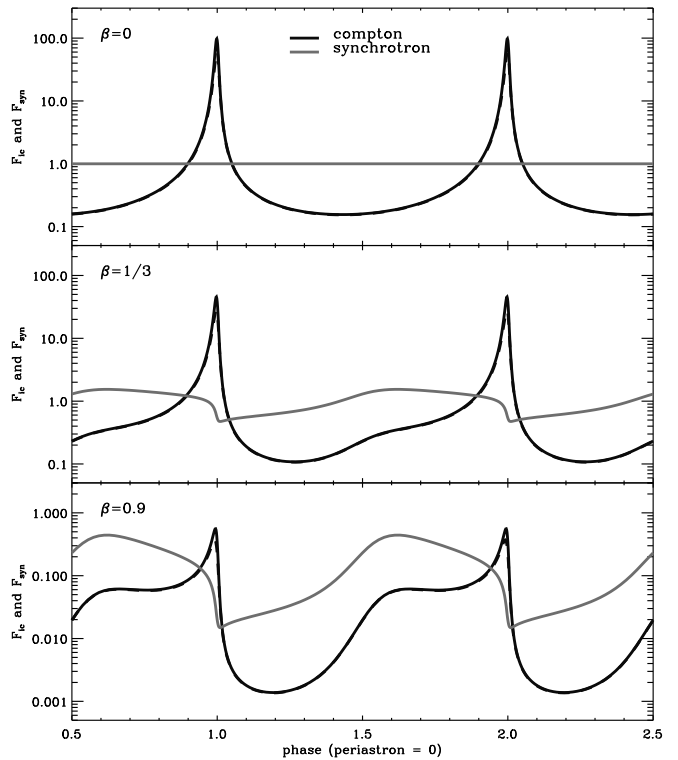}} 
\caption{Doppler-boosted synchrotron ($F_{\rm syn}$, Eq.~\ref{fsyn}) and inverse Compton ($F_{\rm ic}$, Eq.~\ref{fic}) intensity variations  for \psrb\ assuming $\beta$=0 (top), 1/3 (middle) and 0.9 (bottom). The direction of the flow is assumed to be tangent to the orbit ($\theta_{\rm flow}\neq0$). Note the logarithmic y-axis scale. Gamma-ray absorption is negligible here.}
\label{fig_boost_c}
\end{figure}

The case of \psrb\ was also explored under the same assumption as \lsi\ (Fig.~\ref{fig_boost_c}). The inclination is relatively low $i=30\degr$ so that $F_{\rm ic}$ is almost symmetric without Doppler corrections (top panel).  Looking at the top two panels, it can be seen that the Doppler corrections have little impact on the overall lightcurve because of the low inclination. The bottom panel shows that high Doppler factors can strongly deboost the overall lightcurve even though the morphology remains roughly the same. There is no obvious relationship between these curves and the (sparse) X-ray or VHE observations. Other variability factors probably dominate in this much wider binary system.

\citet{2008MNRAS.387...63B} carried out relativistic hydrodynamical simulations of a pulsar wind interacting with a stellar wind with the specific case of \psrb\ in mind. They found that the shocked pulsar wind can accelerate from bulk Lorentz factors $\approx 1$ close to the termination shock up to 100 far away. Emission from such highly relativistic flows is not compatible with observations: the emission would be strongly deboosted (bottom panel, Fig.~\ref{fig_boost_c}) except where (and if) the line-of-sight crosses the relativistic beaming angle where it would produce a flare. The observed X-ray and VHE modulations in gamma-ray binaries suggest modest boosting. The X-ray and VHE emission is more likely to originate close to the termination shock where the jump conditions for an unmagnetized relativistic flow give $\beta=1/3$ \citep{Kennel:1984pd}. 

\section{Conclusion}
The X-ray orbital modulation of \ls\ peaks and falls at conjunctions, suggesting that the underlying mechanism is related to the geometry seen by the observer. Phase-dependent Doppler boosting of emission from a mildly relativistic flow provides a viable explanation. The underlying assumption is that the flow direction changes with orbital phase, so that even constant intrinsic emission becomes variable as seen by the observer. The peaks and troughs are at conjunctions for a flow directed radially away from the star, as expected if the emission arises from a shocked pulsar wind confined by the fast stellar wind of its companion  \citep{Dubus:2006lc}. A moderate relativistic speed of $\beta=0.15$ or 1/3 is enough to reproduce the morphology of the observed X-ray lightcurve assuming (resp.) either constant intrinsic emission or the model of \citet{Dubus:2007oq}. Note that these values of $\beta$ allow for quite large values of the opening angles. More detailed calculations assuming a conical geometry for the flow confirmed that the results were unchanged as long as the angular size of the flow is smaller than $1/\Gamma$ (if larger, the modulation is dampened). Reproducing the level of X-ray emission is difficult with a one-zone model as it requires values of the magnetic field that are a factor 3 above current values, leading to cutoff in the VHE specta at energies that are too low. A more complex multi-zone model of the post-shock flow might resolve this discrepancy.

Inverse Compton scattering in the flow of external stellar photons will be modulated differently than intrinsic emission from the flow. In the case of a radial outflow, the external seed photon flux will be deboosted at all phases. However, a flow tangent to an eccentric orbit, as might arise in \lsi\ and \psrb, can lead both to boosts and deboosts in the comoving frame depending on orbital phase and thus give rise to complex modulations. The calculated Doppler corrected emission in \lsi\  peaks in phase with the observed VHE maximum. This is noteworthy since a simple explanation had not yet been proposed for the phase of VHE (and X-ray) maximum in \lsi. This explanation requires that the shocked pulsar wind flows along the orbit, which appears compatible with the radio VLBI images on larger scales shown in \citet{Dhawan:2006kr}.

The present work assumed a pulsar relativistic wind in the orbital plane but microquasar models have also been proposed for both \ls\ and \lsi. In this case, the emission arises from a relativistic jet.  The jet angle to the observer remains constant along the orbit and so do ${\cal D}_{\rm obs}$ and $F_{\rm syn}$. Hence, no orbital modulation of intrinsic (synchrotron) X-ray emission due to Doppler boosting would be expected, apart from the possible impact of  jet precession on timescales longer than the orbital period \citep{2002A&A...385L..10K}.  Doppler boosting in a relativistic jet cannot explain the X-ray modulation in \ls\ or \lsi.  However, unless the electrons are far from the system or the system is seen pole-on, the angle of interaction between photons and electrons  $\mathbf{e_\star}.\mathbf{e_{\rm obs}}$ will change with orbital phase. A modulation in $F_{\rm ic}$ is unavoidable. This variation in inverse Compton emission can explain the orbital modulation seen in high-energy gamma-rays from the microquasar Cygnus X-3 by  {\em Fermi Gamma-ray Space Telescope} \citep{2009Sci...326.1512Fa,2010arXiv1002.3888D}.

\begin{acknowledgements}
We thank T. Takahashi for sharing the data points plotted in Fig. 2. The authors acknowledge support from the European Community via contract ERC-StG-200911.
\end{acknowledgements}

\bibliographystyle{aa}
\bibliography{../BIBLIO}

\begin{thebibliography}{41}
\expandafter\ifx\csname natexlab\endcsname\relax\def\natexlab#1{#1}\fi

\bibitem[{{Abdo} {et~al.}(2009{\natexlab{a}}){Abdo}, {Ackermann}, {Ajello},
  {Atwood}, {Axelsson}, {Baldini}, {Ballet}, {Barbiellini}, {Bastieri},
  {Baughman}, {Bechtol}, {Bellazzini}, {Berenji}, {Blandford}, {Bloom},
  {Bonamente}, {Borgland}, {Bregeon}, {Brez}, {Brigida}, {Bruel}, {Burnett},
  {Caliandro}, {Cameron}, {Caraveo}, {Casandjian}, {Cavazzuti}, {Cecchi}, {{\c
  C}elik}, {Charles}, {Chaty}, {Chekhtman}, {Cheung}, {Chiang}, {Ciprini},
  {Claus}, {Cohen-Tanugi}, {Cominsky}, {Conrad}, {Corbel}, {Corbet}, {Cutini},
  {Dermer}, {de Angelis}, {de Luca}, {de Palma}, {Digel}, {Dormody}, {do Couto
  e Silva}, {Drell}, {Dubois}, {Dubus}, {Dumora}, {Farnier}, {Favuzzi},
  {Fegan}, {Focke}, {Frailis}, {Fukazawa}, {Funk}, {Fusco}, {Gargano},
  {Gasparrini}, {Gehrels}, {Germani}, {Giebels}, {Giglietto}, {Giordano},
  {Glanzman}, {Godfrey}, {Grenier}, {Grondin}, {Grove}, {Guillemot}, {Guiriec},
  {Hanabata}, {Harding}, {Hayashida}, {Hays}, {Hill}, {Hughes},
  {J{\'o}hannesson}, {Johnson}, {Johnson}, {Johnson}, {Johnson}, {Kamae},
  {Katagiri}, {Kataoka}, {Kawai}, {Kerr}, {Kn{\"o}dlseder}, {Kocian}, {Kuehn},
  {Kuss}, {Lande}, {Larsson}, {Latronico}, {Longo}, {Loparco}, {Lott},
  {Lovellette}, {Lubrano}, {Madejski}, {Makeev}, {Marelli}, {Mazziotta},
  {McEnery}, {Meurer}, {Michelson}, {Mitthumsiri}, {Mizuno}, {Monte},
  {Monzani}, {Morselli}, {Moskalenko}, {Murgia}, {Nolan}, {Nuss}, {Ohsugi},
  {Okumura}, {Omodei}, {Orlando}, {Ormes}, {Paneque}, {Panetta}, {Parent},
  {Pelassa}, {Pepe}, {Pesce-Rollins}, {Piron}, {Porter}, {Rain{\`o}}, {Rando},
  {Ray}, {Razzano}, {Rea}, {Reimer}, {Reimer}, {Reposeur}, {Ritz}, {Rochester},
  {Rodriguez}, {Romani}, {Ryde}, {Sadrozinski}, {Sanchez}, {Sander}, {Saz
  Parkinson}, {Scargle}, {Sgr{\`o}}, {Shaw}, {Sierpowska-Bartosik}, {Siskind},
  {Smith}, {Smith}, {Spandre}, {Spinelli}, {Striani}, {Strickman}, {Suson},
  {Tajima}, {Takahashi}, {Takahashi}, {Tanaka}, {Thayer}, {Thayer}, {Thompson},
  {Tibaldo}, {Torres}, {Tosti}, {Tramacere}, {Uchiyama}, {Usher}, {Vasileiou},
  {Vilchez}, {Vitale}, {Waite}, {Wang}, {Winer}, {Wood}, {Ylinen}, \&
  {Ziegler}}]{2009ApJ...701L.123A}
{Abdo}, A.~A., {Ackermann}, M., {Ajello}, M., {et~al.} 2009{\natexlab{a}},
  \apjl, 701, L123

\bibitem[{{Abdo} {et~al.}(2009{\natexlab{b}}){Abdo}, {Ackermann}, {Ajello},
  {Atwood}, {Axelsson}, {Baldini}, {Ballet}, {Barbiellini}, {Bastieri},
  {Baughman}, {Bechtol}, {Bellazzini}, {Berenji}, {Blandford}, {Bloom},
  {Bonamente}, {Borgland}, {Bregeon}, {Brez}, {Brigida}, {Bruel}, {Burnett},
  {Buson}, {Caliandro}, {Cameron}, {Caraveo}, {Casandjian}, {Cavazzuti},
  {Cecchi}, {{\c C}elik}, {Chaty}, {Chekhtman}, {Cheung}, {Chiang}, {Ciprini},
  {Claus}, {Cohen-Tanugi}, {Cominsky}, {Conrad}, {Corbel}, {Corbet}, {Cutini},
  {Dermer}, {de Angelis}, {de Palma}, {Digel}, {Silva}, {Drell}, {Dubois},
  {Dubus}, {Dumora}, {Farnier}, {Favuzzi}, {Fegan}, {Focke}, {Fortin},
  {Frailis}, {Fukazawa}, {Funk}, {Fusco}, {Gargano}, {Gasparrini}, {Gehrels},
  {Germani}, {Giebels}, {Giglietto}, {Giordano}, {Glanzman}, {Godfrey},
  {Grenier}, {Grondin}, {Grove}, {Guillemot}, {Guiriec}, {Hanabata}, {Harding},
  {Hayashida}, {Hays}, {Hill}, {Horan}, {Hughes}, {Jackson}, {J{\'o}hannesson},
  {Johnson}, {Johnson}, {Johnson}, {Kamae}, {Katagiri}, {Kataoka}, {Kawai},
  {Kerr}, {Kn{\"o}dlseder}, {Kocian}, {Kuehn}, {Kuss}, {Lande}, {Larsson},
  {Latronico}, {Lemoine-Goumard}, {Longo}, {Loparco}, {Lott}, {Lovellette},
  {Lubrano}, {Madejski}, {Makeev}, {Marelli}, {Mazziotta}, {McEnery}, {Meurer},
  {Michelson}, {Mitthumsiri}, {Mizuno}, {Moiseev}, {Monte}, {Monzani},
  {Morselli}, {Moskalenko}, {Murgia}, {Nolan}, {Norris}, {Nuss}, {Ohsugi},
  {Omodei}, {Orlando}, {Ormes}, {Ozaki}, {Paneque}, {Panetta}, {Parent},
  {Pelassa}, {Pepe}, {Pesce-Rollins}, {Piron}, {Porter}, {Rain{\`o}}, {Rando},
  {Ray}, {Razzano}, {Rea}, {Reimer}, {Reimer}, {Reposeur}, {Ritz}, {Rochester},
  {Rodriguez}, {Romani}, {Roth}, {Ryde}, {Sadrozinski}, {Sanchez}, {Sander},
  {Saz Parkinson}, {Scargle}, {Sgr{\`o}}, {Sierpowska-Bartosik}, {Siskind},
  {Smith}, {Smith}, {Spandre}, {Spinelli}, {Strickman}, {Suson}, {Tajima},
  {Takahashi}, {Takahashi}, {Tanaka}, {Tanaka}, {Thayer}, {Thompson},
  {Tibaldo}, {Torres}, {Tosti}, {Tramacere}, {Uchiyama}, {Usher}, {Vasileiou},
  {Venter}, {Vilchez}, {Vitale}, {Waite}, {Wallace}, {Wang}, {Winer}, {Wood},
  {Ylinen}, \& {Ziegler}}]{2009ApJ...706L..56A}
{Abdo}, A.~A., {Ackermann}, M., {Ajello}, M., {et~al.} 2009{\natexlab{b}},
  \apjl, 706, L56

\bibitem[{{Abdo} {et~al.}(2009{\natexlab{c}}){Abdo}, {Ackermann}, {Ajello},
  {Axelsson}, {Baldini}, {Ballet}, {Barbiellini}, {Bastieri}, {Baughman},
  {Bechtol}, {Bellazzini}, {Berenji}, {Blandford}, {Bloom}, {Bonamente},
  {Borgland}, {Brez}, {Brigida}, {Bruel}, {Burnett}, {Buson}, {Caliandro},
  {Cameron}, {Caraveo}, {Casandjian}, {Cecchi}, {{\c C}elik}, {Chaty},
  {Cheung}, {Chiang}, {Ciprini}, {Claus}, {Cohen-Tanugi}, {Cominsky}, {Conrad},
  {Corbel}, {Corbet}, {Dermer}, {de Palma}, {Digel}, {do Couto e Silva},
  {Drell}, {Dubois}, {Dubus}, {Dumora}, {Farnier}, {Favuzzi}, {Fegan}, {Focke},
  {Fortin}, {Frailis}, {Fusco}, {Gargano}, {Gehrels}, {Germani}, {Giavitto},
  {Giebels}, {Giglietto}, {Giordano}, {Glanzman}, {Godfrey}, {Grenier},
  {Grondin}, {Grove}, {Guillemot}, {Guiriec}, {Hanabata}, {Harding},
  {Hayashida}, {Hays}, {Hill}, {Hjalmarsdotter}, {Horan}, {Hughes}, {Jackson},
  {J{\'o}hannesson}, {Johnson}, {Johnson}, {Johnson}, {Kamae}, {Katagiri},
  {Kawai}, {Kerr}, {Kn{\"o}dlseder}, {Kocian}, {Koerding}, {Kuss}, {Lande},
  {Latronico}, {Lemoine-Goumard}, {Longo}, {Loparco}, {Lott}, {Lovellette},
  {Lubrano}, {Madejski}, {Makeev}, {Marchand}, {Marelli}, {Max-Moerbeck},
  {Mazziotta}, {McColl}, {McEnery}, {Meurer}, {Michelson}, {Migliari},
  {Mitthumsiri}, {Mizuno}, {Monte}, {Monzani}, {Morselli}, {Moskalenko},
  {Murgia}, {Nolan}, {Norris}, {Nuss}, {Ohsugi}, {Omodei}, {Ong}, {Ormes},
  {Paneque}, {Parent}, {Pelassa}, {Pepe}, {Pesce-Rollins}, {Piron}, {Pooley},
  {Porter}, {Pottschmidt}, {Rain{\`o}}, {Rando}, {Ray}, {Razzano}, {Rea},
  {Readhead}, {Reimer}, {Reimer}, {Richards}, {Rochester}, {Rodriguez},
  {Rodriguez}, {Romani}, {Ryde}, {Sadrozinski}, {Sander}, {Saz Parkinson},
  {Sgr{\`o}}, {Siskind}, {Smith}, {Smith}, {Spinelli}, {Starck}, {Stevenson},
  {Strickman}, {Suson}, {Takahashi}, {Tanaka}, {Thayer}, {Thompson}, {Tibaldo},
  {Tomsick}, {Torres}, {Tosti}, {Tramacere}, {Uchiyama}, {Usher}, {Vasileiou},
  {Vilchez}, {Vitale}, {Waite}, {Wang}, {Wilms}, {Winer}, {Wood}, {Ylinen}, \&
  {Ziegler}}]{2009Sci...326.1512Fa}
{Abdo}, A.~A., {Ackermann}, M., {Ajello}, M., {et~al.} 2009{\natexlab{c}},
  Science, 326, 1512

\bibitem[{{Acciari} {et~al.}(2009){Acciari}, {Aliu}, {Arlen}, {Bautista},
  {Beilicke}, {Benbow}, {B{\"o}ttcher}, {Bradbury}, {Bugaev}, {Butt}, {Butt},
  {Byrum}, {Cannon}, {Cesarini}, {Chow}, {Ciupik}, {Cogan}, {Colin}, {Cui},
  {Daniel}, {Dickherber}, {Ergin}, {Falcone}, {Fegan}, {Finley}, {Fortin},
  {Fortson}, {Furniss}, {Gall}, {Gillanders}, {Grube}, {Guenette}, {Gyuk},
  {Hanna}, {Hays}, {Holder}, {Horan}, {Hui}, {Humensky}, {Kaaret}, {Karlsson},
  {Kieda}, {Kildea}, {Konopelko}, {Krawczynski}, {Krennrich}, {Lang},
  {LeBohec}, {Maier}, {McCann}, {McCutcheon}, {Millis}, {Moriarty},
  {Mukherjee}, {Nagai}, {Ong}, {Otte}, {Pandel}, {Perkins}, {Perkins}, {Pohl},
  {Quinn}, {Ragan}, {Reyes}, {Reynolds}, {Roache}, {Joachim Rose},
  {Schroedter}, {Sembroski}, {Smith}, {Steele}, {Stroh}, {Swordy}, {Theiling},
  {Toner}, {Varlotta}, {Vassiliev}, {Wagner}, {Wakely}, {Ward}, {Weekes},
  {Weinstein}, {White}, {Williams}, {Wissel}, {Wood}, \&
  {Zitzer}}]{2009ApJ...700.1034A}
{Acciari}, V.~A., {Aliu}, E., {Arlen}, T., {et~al.} 2009, \apj, 700, 1034

\bibitem[{{Acciari} {et~al.}(2008){Acciari}, {Beilicke}, {Blaylock},
  {Bradbury}, {Buckley}, {Bugaev}, {Butt}, {Byrum}, {Celik}, {Cesarini},
  {Ciupik}, {Chow}, {Cogan}, {Colin}, {Cui}, {Daniel}, {Duke}, {Ergin},
  {Falcone}, {Fegan}, {Finley}, {Fortin}, {Fortson}, {Gall}, {Gibbs},
  {Gillanders}, {Grube}, {Guenette}, {Hanna}, {Hays}, {Holder}, {Horan},
  {Hughes}, {Hui}, {Humensky}, {Kaaret}, {Kieda}, {Kildea}, {Konopelko},
  {Krawczynski}, {Krennrich}, {Lang}, {LeBohec}, {Lee}, {Maier}, {McCann},
  {McCutcheon}, {Millis}, {Moriarty}, {Mukherjee}, {Nagai}, {Ong}, {Pandel},
  {Perkins}, {Pizlo}, {Pohl}, {Quinn}, {Ragan}, {Reynolds}, {Rose},
  {Schroedter}, {Sembroski}, {Smith}, {Steele}, {Swordy}, {Toner}, {Valcarcel},
  {Vassiliev}, {Wagner}, {Wakely}, {Ward}, {Weekes}, {Weinstein}, {White},
  {Williams}, {Wissel}, {Wood}, \& {Zitzer}}]{Acciari:2008vf}
{Acciari}, V.~A., {Beilicke}, M., {Blaylock}, G., {et~al.} 2008, \apj, 679,
  1427

\bibitem[{{Aharonian} {et~al.}(2005{\natexlab{a}}){Aharonian}, {Akhperjanian},
  {Aye}, {Bazer-Bachi}, {Beilicke}, {Benbow}, {Berge}, {Berghaus},
  {Bernl{\"o}hr}, {Boisson}, {Bolz}, {Borrel}, {Braun}, {Breitling}, {Brown},
  {Gordo}, {Chadwick}, {Chounet}, {Cornils}, {Costamante}, {Degrange},
  {Dickinson}, {Djannati-Ata{\"\i}}, {Drury}, {Dubus}, {Emmanoulopoulos},
  {Espigat}, {Feinstein}, {Fleury}, {Fontaine}, {Fuchs}, {Funk}, {Gallant},
  {Giebels}, {Gillessen}, {Glicenstein}, {Goret}, {Hadjichristidis}, {Hauser},
  {Heinzelmann}, {Henri}, {Hermann}, {Hinton}, {Hofmann}, {Holleran}, {Horns},
  {Jacholkowska}, {de Jager}, {Kh{\'e}lifi}, {Komin}, {Konopelko}, {Latham},
  {Le Gallou}, {Lemi{\`e}re}, {Lemoine-Goumard}, {Leroy}, {Lohse}, {Marcowith},
  {Martin}, {Martineau-Huynh}, {Masterson}, {McComb}, {de Naurois}, {Nolan},
  {Noutsos}, {Orford}, {Osborne}, {Ouchrif}, {Panter}, {Pelletier}, {Pita},
  {P{\"u}hlhofer}, {Punch}, {Raubenheimer}, {Raue}, {Raux}, {Rayner}, {Reimer},
  {Reimer}, {Ripken}, {Rob}, {Rolland}, {Rowell}, {Sahakian}, {Saug{\'e}},
  {Schlenker}, {Schlickeiser}, {Schuster}, {Schwanke}, {Siewert}, {Sol},
  {Spangler}, {Steenkamp}, {Stegmann}, {Tavernet}, {Terrier}, {Th{\'e}oret},
  {Tluczykont}, {Vasileiadis}, {Venter}, {Vincent}, {V{\"o}lk}, \&
  {Wagner}}]{Aharonian:2005nj}
{Aharonian}, F., {Akhperjanian}, A.~G., {Aye}, K.-M., {et~al.}
  2005{\natexlab{a}}, Science, 309, 746

\bibitem[{{Aharonian} {et~al.}(2005{\natexlab{b}}){Aharonian}, {Akhperjanian},
  {Aye}, {Bazer-Bachi}, {Beilicke}, {Benbow}, {Berge}, {Berghaus},
  {Bernl{\"o}hr}, {Boisson}, {Bolz}, {Braun}, {Breitling}, {Brown}, {Bussons
  Gordo}, {Chadwick}, {Chounet}, {Cornils}, {Costamante}, {Degrange},
  {Djannati-Ata{\"\i}}, {O'C.~Drury}, {Dubus}, {Emmanoulopoulos}, {Espigat},
  {Feinstein}, {Fleury}, {Fontaine}, {Fuchs}, {Funk}, {Gallant}, {Giebels},
  {Gillessen}, {Glicenstein}, {Goret}, {Hadjichristidis}, {Hauser},
  {Heinzelmann}, {Henri}, {Hermann}, {Hinton}, {Hofmann}, {Holleran}, {Horns},
  {de Jager}, {Johnston}, {Kh{\'e}lifi}, {Kirk}, {Komin}, {Konopelko},
  {Latham}, {Le Gallou}, {Lemi{\`e}re}, {Lemoine-Goumard}, {Leroy},
  {Martineau-Huynh}, {Lohse}, {Marcowith}, {Masterson}, {McComb}, {de Naurois},
  {Nolan}, {Noutsos}, {Orford}, {Osborne}, {Ouchrif}, {Panter}, {Pelletier},
  {Pita}, {P{\"u}hlhofer}, {Punch}, {Raubenheimer}, {Raue}, {Raux}, {Rayner},
  {Redondo}, {Reimer}, {Reimer}, {Ripken}, {Rob}, {Rolland}, {Rowell},
  {Sahakian}, {Saug{\'e}}, {Schlenker}, {Schlickeiser}, {Schuster}, {Schwanke},
  {Siewert}, {Skj{\ae}raasen}, {Sol}, {Steenkamp}, {Stegmann}, {Tavernet},
  {Terrier}, {Th{\'e}oret}, {Tluczykont}, {Vasileiadis}, {Venter}, {Vincent},
  {V{\"o}lk}, \& {Wagner}}]{Aharonian:2005br}
{Aharonian}, F., {Akhperjanian}, A.~G., {Aye}, K.-M., {et~al.}
  2005{\natexlab{b}}, \aap, 442, 1

\bibitem[{{Aharonian} {et~al.}(2006){Aharonian}, {Akhperjanian}, {Bazer-Bachi},
  {Beilicke}, {Benbow}, {Berge}, {Bernl{\"o}hr}, {Boisson}, {Bolz}, {Borrel},
  {Braun}, {Brown}, {B{\"u}hler}, {B{\"u}sching}, {Carrigan}, {Chadwick},
  {Chounet}, {Cornils}, {Costamante}, {Degrange}, {Dickinson},
  {Djannati-Ata{\"\i}}, {O'C.~Drury}, {Dubus}, {Egberts}, {Emmanoulopoulos},
  {Espigat}, {Feinstein}, {Ferrero}, {Fiasson}, {Fontaine}, {Funk}, {Funk},
  {F{\"u}{\ss}ling}, {Gallant}, {Giebels}, {Glicenstein}, {Goret},
  {Hadjichristidis}, {Hauser}, {Hauser}, {Heinzelmann}, {Henri}, {Hermann},
  {Hinton}, {Hoffmann}, {Hofmann}, {Holleran}, {Horns}, {Jacholkowska}, {de
  Jager}, {Kendziorra}, {Kh{\'e}lifi}, {Komin}, {Konopelko}, {Kosack},
  {Latham}, {Le Gallou}, {Lemi{\`e}re}, {Lemoine-Goumard}, {Lohse}, {Martin},
  {Martineau-Huynh}, {Marcowith}, {Masterson}, {Maurin}, {McComb}, {Moulin},
  {de Naurois}, {Nedbal}, {Nolan}, {Noutsos}, {Orford}, {Osborne}, {Ouchrif},
  {Panter}, {Pelletier}, {Pita}, {P{\"u}hlhofer}, {Punch}, {Raubenheimer},
  {Raue}, {Rayner}, {Reimer}, {Reimer}, {Ripken}, {Rob}, {Rolland}, {Rowell},
  {Sahakian}, {Santangelo}, {Saug{\'e}}, {Schlenker}, {Schlickeiser},
  {Schr{\"o}der}, {Schwanke}, {Schwarzburg}, {Shalchi}, {Sol}, {Spangler},
  {Spanier}, {Steenkamp}, {Stegmann}, {Superina}, {Tavernet}, {Terrier},
  {Tluczykont}, {van Eldik}, {Vasileiadis}, {Venter}, {Vincent}, {V{\"o}lk},
  {Wagner}, \& {Ward}}]{Aharonian:2006qw}
{Aharonian}, F., {Akhperjanian}, A.~G., {Bazer-Bachi}, A.~R., {et~al.} 2006,
  \aap, 460, 743

\bibitem[{{Albert} {et~al.}(2009){Albert}, {Aliu}, {Anderhub}, {Antonelli},
  {Antoranz}, {Backes}, {Baixeras}, {Barrio}, {Bartko}, {Bastieri}, {Becker},
  {Bednarek}, {Berger}, {Bernardini}, {Bigongiari}, {Biland}, {Bock},
  {Bonnoli}, {Bordas}, {Bosch-Ramon}, {Bretz}, {Britvitch}, {Camara},
  {Carmona}, {Chilingarian}, {Commichau}, {Contreras}, {Cortina}, {Costado},
  {Covino}, {Curtef}, {Dazzi}, {DeAngelis}, {DeCea del Pozo}, {de los Reyes},
  {DeLotto}, {DeMaria}, {DeSabata}, {Delgado Mendez}, {Dominguez}, {Dorner},
  {Doro}, {Errando}, {Fagiolini}, {Ferenc}, {Fern{\'a}ndez}, {Firpo},
  {Fonseca}, {Font}, {Galante}, {Garc{\'{\i}}a L{\'o}pez}, {Garczarczyk},
  {Gaug}, {Goebel}, {Hayashida}, {Herrero}, {H{\"o}hne}, {Hose}, {Hsu},
  {Huber}, {Jogler}, {Kranich}, {La Barbera}, {Laille}, {Leonardo}, {Lindfors},
  {Lombardi}, {Longo}, {L{\'o}pez}, {Lorenz}, {Majumdar}, {Maneva},
  {Mankuzhiyil}, {Mannheim}, {Maraschi}, {Mariotti}, {Mart{\'{\i}}nez},
  {Mazin}, {Meucci}, {Meyer}, {Miranda}, {Mirzoyan}, {Mizobuchi}, {Moles},
  {Moralejo}, {Nieto}, {Nilsson}, {Ninkovic}, {Otte}, {Oya}, {Panniello},
  {Paoletti}, {Paredes}, {Pasanen}, {Pascoli}, {Pauss}, {Pegna},
  {Perez-Torres}, {Persic}, {Peruzzo}, {Piccioli}, {Prada}, {Prandini},
  {Puchades}, {Raymers}, {Rhode}, {Rib{\'o}}, {Rico}, {Rissi}, {Robert},
  {R{\"u}gamer}, {Saggion}, {Saito}, {Salvati}, {Sanchez-Conde}, {Sartori},
  {Satalecka}, {Scalzotto}, {Scapin}, {Schmitt}, {Schweizer}, {Shayduk},
  {Shinozaki}, {Shore}, {Sidro}, {Sierpowska-Bartosik}, {Sillanp{\"a}{\"a}},
  {Sobczynska}, {Spanier}, {Stamerra}, {Stark}, {Takalo}, {Tavecchio},
  {Temnikov}, {Tescaro}, {Teshima}, {Tluczykont}, {Torres}, {Turini}, {Vankov},
  {Venturini}, {Vitale}, {Wagner}, {Wittek}, {Zabalza}, {Zandanel}, {Zanin}, \&
  {Zapatero}}]{2009ApJ...693..303A}
{Albert}, J., {Aliu}, E., {Anderhub}, H., {et~al.} 2009, \apj, 693, 303

\bibitem[{{Albert} {et~al.}(2006){Albert}, {Aliu}, {Anderhub}, {Antoranz},
  {Armada}, {Asensio}, {Baixeras}, {Barrio}, {Bartelt}, {Bartko}, {Bastieri},
  {Bavikadi}, {Bednarek}, {Berger}, {Bigongiari}, {Biland}, {Bisesi}, {Bock},
  {Bordas}, {Bosch-Ramon}, {Bretz}, {Britvitch}, {Camara}, {Carmona},
  {Chilingarian}, {Ciprini}, {Coarasa}, {Commichau}, {Contreras}, {Cortina},
  {Curtef}, {Danielyan}, {Dazzi}, {De Angelis}, {de los Reyes}, {De Lotto},
  {Domingo-Santamar{\'{\i}}a}, {Dorner}, {Doro}, {Errando}, {Fagiolini},
  {Ferenc}, {Fern{\'a}ndez}, {Firpo}, {Flix}, {Fonseca}, {Font}, {Fuchs},
  {Galante}, {Garczarczyk}, {Gaug}, {Giller}, {Goebel}, {Hakobyan},
  {Hayashida}, {Hengstebeck}, {H{\"o}hne}, {Hose}, {Hsu}, {Isar}, {Jacon},
  {Kalekin}, {Kosyra}, {Kranich}, {Laatiaoui}, {Laille}, {Lenisa}, {Liebing},
  {Lindfors}, {Lombardi}, {Longo}, {L{\'o}pez}, {L{\'o}pez}, {Lorenz},
  {Lucarelli}, {Majumdar}, {Maneva}, {Mannheim}, {Mansutti}, {Mariotti},
  {Mart{\'{\i}}nez}, {Mase}, {Mazin}, {Merck}, {Meucci}, {Meyer}, {Miranda},
  {Mirzoyan}, {Mizobuchi}, {Moralejo}, {Nilsson}, {O{\~n}a-Wilhelmi},
  {Ordu{\~n}a}, {Otte}, {Oya}, {Paneque}, {Paoletti}, {Paredes}, {Pasanen},
  {Pascoli}, {Pauss}, {Pavel}, {Pegna}, {Persic}, {Peruzzo}, {Piccioli},
  {Poller}, {Pooley}, {Prandini}, {Raymers}, {Rhode}, {Rib{\'o}}, {Rico},
  {Riegel}, {Rissi}, {Robert}, {Romero}, {R{\"u}gamer}, {Saggion},
  {S{\'a}nchez}, {Sartori}, {Scalzotto}, {Scapin}, {Schmitt}, {Schweizer},
  {Shayduk}, {Shinozaki}, {Shore}, {Sidro}, {Sillanp{\"a}{\"a}}, {Sobczynska},
  {Stamerra}, {Stark}, {Takalo}, {Temnikov}, {Tescaro}, {Teshima}, {Tonello},
  {Torres}, {Torres}, {Turini}, {Vankov}, {Vitale}, {Wagner}, {Wibig},
  {Wittek}, {Zanin}, \& {Zapatero}}]{Albert:2006wi}
{Albert}, J., {Aliu}, E., {Anderhub}, H., {et~al.} 2006, Science, 312, 1771

\bibitem[{{Anderhub} {et~al.}(2009){Anderhub}, {Antonelli}, {Antoranz},
  {Backes}, {Baixeras}, {Balestra}, {Barrio}, {Bastieri}, {Becerra
  Gonz{\'a}lez}, {Becker}, {Bednarek}, {Berger}, {Bernardini}, {Biland},
  {Blanch Bigas}, {Bock}, {Bonnoli}, {Bordas}, {Borla Tridon}, {Bosch-Ramon},
  {Bose}, {Braun}, {Bretz}, {Britzger}, {Camara}, {Carmona}, {Carosi}, {Colin},
  {Commichau}, {Contreras}, {Cortina}, {Costado}, {Covino}, {Dazzi}, {De
  Angelis}, {de Cea del Pozo}, {De los Reyes}, {De Lotto}, {De Maria}, {De
  Sabata}, {Delgado Mendez}, {Dom{\'{\i}}nguez}, {Dominis Prester}, {Dorner},
  {Doro}, {Elsaesser}, {Errando}, {Ferenc}, {Fern{\'a}ndez}, {Firpo},
  {Fonseca}, {Font}, {Galante}, {Garc{\'{\i}}a L{\'o}pez}, {Garczarczyk},
  {Gaug}, {Godinovic}, {Goebel}, {Hadasch}, {Herrero}, {Hildebrand},
  {H{\"o}hne-M{\"o}nch}, {Hose}, {Hrupec}, {Hsu}, {Jogler}, {Klepser},
  {Kranich}, {La Barbera}, {Laille}, {Leonardo}, {Lindfors}, {Lombardi},
  {Longo}, {L{\'o}pez}, {Lorenz}, {Majumdar}, {Maneva}, {Mankuzhiyil},
  {Mannheim}, {Maraschi}, {Mariotti}, {Mart{\'{\i}}nez}, {Mazin}, {Meucci},
  {Miranda}, {Mirzoyan}, {Miyamoto}, {Mold{\'o}n}, {Moles}, {Moralejo},
  {Nieto}, {Nilsson}, {Ninkovic}, {Orito}, {Oya}, {Paoletti}, {Paredes},
  {Pasanen}, {Pascoli}, {Pauss}, {Pegna}, {Perez-Torres}, {Persic}, {Peruzzo},
  {Prada}, {Prandini}, {Puchades}, {Puljak}, {Reichardt}, {Rhode}, {Rib{\'o}},
  {Rico}, {Rissi}, {Robert}, {R{\"u}gamer}, {Saggion}, {Saito}, {Salvati},
  {S{\'a}nchez-Conde}, {Satalecka}, {Scalzotto}, {Scapin}, {Schweizer},
  {Shayduk}, {Shore}, {Sidro}, {Sierpowska-Bartosik}, {Sillanp{\"a}{\"a}},
  {Sitarek}, {Sobczynska}, {Spanier}, {Spiro}, {Stamerra}, {Stark}, {Suric},
  {Takalo}, {Tavecchio}, {Temnikov}, {Tescaro}, {Teshima}, {Torres}, {Turini},
  {Vankov}, {Wagner}, {Zabalza}, {Zandanel}, {Zanin}, {Zapatero}, {The MAGIC
  Collaboration}, {Falcone}, {Vetere}, {Gehrels}, {Trushkin}, {Dhawan}, \&
  {Reig}}]{2009ApJ...706L..27A}
{Anderhub}, H., {Antonelli}, L.~A., {Antoranz}, P., {et~al.} 2009, \apjl, 706,
  L27

\bibitem[{{Aragona} {et~al.}(2009){Aragona}, {McSwain}, {Grundstrom}, {Marsh},
  {Roettenbacher}, {Hessler}, {Boyajian}, \& {Ray}}]{2009ApJ...698..514A}
{Aragona}, C., {McSwain}, M.~V., {Grundstrom}, E.~D., {et~al.} 2009, \apj, 698,
  514

\bibitem[{{Arons} \& {Tavani}(1993)}]{1993ApJ...403..249A}
{Arons}, J. \& {Tavani}, M. 1993, \apj, 403, 249

\bibitem[{{Bignami} {et~al.}(1977){Bignami}, {Maraschi}, \&
  {Treves}}]{1977A&A....55..155B}
{Bignami}, G.~F., {Maraschi}, L., \& {Treves}, A. 1977, \aap, 55, 155

\bibitem[{{Bogovalov} {et~al.}(2008){Bogovalov}, {Khangulyan}, {Koldoba},
  {Ustyugova}, \& {Aharonian}}]{2008MNRAS.387...63B}
{Bogovalov}, S.~V., {Khangulyan}, D.~V., {Koldoba}, A.~V., {Ustyugova}, G.~V.,
  \& {Aharonian}, F.~A. 2008, \mnras, 387, 63

\bibitem[{{Bosch-Ramon} {et~al.}(2007){Bosch-Ramon}, {Motch}, {Rib{\'o}},
  {Lopes de Oliveira}, {Janot-Pacheco}, {Negueruela}, {Paredes}, \&
  {Martocchia}}]{Bosch-Ramon:2007fq}
{Bosch-Ramon}, V., {Motch}, C., {Rib{\'o}}, M., {et~al.} 2007, \aap, 473, 545

\bibitem[{{Bosch-Ramon} {et~al.}(2005){Bosch-Ramon}, {Paredes}, {Rib{\'o}},
  {Miller}, {Reig}, \& {Mart{\'{\i}}}}]{Bosch-Ramon:2005zc}
{Bosch-Ramon}, V., {Paredes}, J.~M., {Rib{\'o}}, M., {et~al.} 2005, \apj, 628,
  388

\bibitem[{{Chernyakova} {et~al.}(2009){Chernyakova}, {Neronov}, {Aharonian},
  {Uchiyama}, \& {Takahashi}}]{2009MNRAS.397.2123C}
{Chernyakova}, M., {Neronov}, A., {Aharonian}, F., {Uchiyama}, Y., \&
  {Takahashi}, T. 2009, \mnras, 397, 2123

\bibitem[{{Chernyakova} {et~al.}(2006){Chernyakova}, {Neronov}, \&
  {Walter}}]{Chernyakova:2006cu}
{Chernyakova}, M., {Neronov}, A., \& {Walter}, R. 2006, \mnras, 372, 1585

\bibitem[{{Dermer} \& {Schlickeiser}(1993)}]{1993ApJ...416..458D}
{Dermer}, C.~D. \& {Schlickeiser}, R. 1993, \apj, 416, 458

\bibitem[{{Dermer} {et~al.}(1992){Dermer}, {Schlickeiser}, \&
  {Mastichiadis}}]{1992A&A...256L..27D}
{Dermer}, C.~D., {Schlickeiser}, R., \& {Mastichiadis}, A. 1992, \aap, 256, L27

\bibitem[{{Dhawan} {et~al.}(2006){Dhawan}, {Mioduszewski}, \&
  {Rupen}}]{Dhawan:2006kr}
{Dhawan}, V., {Mioduszewski}, A., \& {Rupen}, M. 2006, in VI Microquasar
  Workshop: Microquasars and Beyond, Vol. MQW6 (Proceedings of Science), 52

\bibitem[{{Dubus}(2006)}]{Dubus:2006lc}
{Dubus}, G. 2006, \aap, 456, 801

\bibitem[{{Dubus} {et~al.}(2008){Dubus}, {Cerutti}, \& {Henri}}]{Dubus:2007oq}
{Dubus}, G., {Cerutti}, B., \& {Henri}, G. 2008, \aap, 477, 691

\bibitem[{{Dubus} {et~al.}(2010){Dubus}, {Cerutti}, \&
  {Henri}}]{2010arXiv1002.3888D}
{Dubus}, G., {Cerutti}, B., \& {Henri}, G. 2010, \mnras, accepted,
  arXiv1002.3888D

\bibitem[{{Georganopoulos} {et~al.}(2001){Georganopoulos}, {Kirk}, \&
  {Mastichiadis}}]{2001ApJ...561..111G}
{Georganopoulos}, M., {Kirk}, J.~G., \& {Mastichiadis}, A. 2001, \apj, 561, 111

\bibitem[{{Hinton} {et~al.}(2009){Hinton}, {Skilton}, {Funk}, {Brucker},
  {Aharonian}, {Dubus}, {Fiasson}, {Gallant}, {Hofmann}, {Marcowith}, \&
  {Reimer}}]{Hinton:2008eg}
{Hinton}, J.~A., {Skilton}, J.~L., {Funk}, S., {et~al.} 2009, \apjl, 690, L101

\bibitem[{{Hoffmann} {et~al.}(2009){Hoffmann}, {Klochkov}, {Santangelo},
  {Horns}, {Segreto}, {Staubert}, \& {Puehlhofer}}]{Hoffmann:2008ys}
{Hoffmann}, A.~D., {Klochkov}, D., {Santangelo}, A., {et~al.} 2009, \aap, 494,
  L37

\bibitem[{{Huang} \& {Becker}(2007)}]{2007A&A...463L...5H}
{Huang}, H.~H. \& {Becker}, W. 2007, \aap, 463, L5

\bibitem[{{Kaufman Bernad{\'o}} {et~al.}(2002){Kaufman Bernad{\'o}}, {Romero},
  \& {Mirabel}}]{2002A&A...385L..10K}
{Kaufman Bernad{\'o}}, M.~M., {Romero}, G.~E., \& {Mirabel}, I.~F. 2002, \aap,
  385, L10

\bibitem[{{Kennel} \& {Coroniti}(1984)}]{Kennel:1984pd}
{Kennel}, C.~F. \& {Coroniti}, F.~V. 1984, \apj, 283, 694

\bibitem[{{Kishishita} {et~al.}(2009){Kishishita}, {Tanaka}, {Uchiyama}, \&
  {Takahashi}}]{2009ApJ...697L...1K}
{Kishishita}, T., {Tanaka}, T., {Uchiyama}, Y., \& {Takahashi}, T. 2009, \apjl,
  697, L1

\bibitem[{{Manchester} {et~al.}(1995){Manchester}, {Johnston}, {Lyne},
  {D'Amico}, {Bailes}, \& {Nicastro}}]{Manchester:1995ck}
{Manchester}, R.~N., {Johnston}, S., {Lyne}, A.~G., {et~al.} 1995, \apjl, 445,
  L137

\bibitem[{{Maraschi} \& {Treves}(1981)}]{Maraschi:1981nj}
{Maraschi}, L. \& {Treves}, A. 1981, \mnras, 194, 1P

\bibitem[{{Martocchia} {et~al.}(2005){Martocchia}, {Motch}, \&
  {Negueruela}}]{Martocchia:2005kq}
{Martocchia}, A., {Motch}, C., \& {Negueruela}, I. 2005, \aap, 430, 245

\bibitem[{{Motch} {et~al.}(1997){Motch}, {Haberl}, {Dennerl}, {Pakull}, \&
  {Janot-Pacheco}}]{Motch:1997qk}
{Motch}, C., {Haberl}, F., {Dennerl}, K., {Pakull}, M., \& {Janot-Pacheco}, E.
  1997, \aap, 323, 853

\bibitem[{{Pelling} {et~al.}(1987){Pelling}, {Paciesas}, {Peterson},
  {Makishima}, {Oda}, {Ogawara}, \& {Miyamoto}}]{1987ApJ...319..416P}
{Pelling}, R.~M., {Paciesas}, W.~S., {Peterson}, L.~E., {et~al.} 1987, \apj,
  319, 416

\bibitem[{{Reig} {et~al.}(2003){Reig}, {Rib{\'o}}, {Paredes}, \&
  {Mart{\'{\i}}}}]{Reig:2003eq}
{Reig}, P., {Rib{\'o}}, M., {Paredes}, J.~M., \& {Mart{\'{\i}}}, J. 2003, \aap,
  405, 285

\bibitem[{{Rib{\'o}} {et~al.}(1999){Rib{\'o}}, {Reig}, {Mart{\'{\i}}}, \&
  {Paredes}}]{Ribo:1999qo}
{Rib{\'o}}, M., {Reig}, P., {Mart{\'{\i}}}, J., \& {Paredes}, J.~M. 1999, \aap,
  347, 518

\bibitem[{{Takahashi} {et~al.}(2009){Takahashi}, {Kishishita}, {Uchiyama},
  {Tanaka}, {Yamaoka}, {Khangulyan}, {Aharonian}, {Bosch-Ramon}, \&
  {Hinton}}]{Takahashi:2008vu}
{Takahashi}, T., {Kishishita}, T., {Uchiyama}, Y., {et~al.} 2009, \apj, 697,
  592

\bibitem[{{Tavani} {et~al.}(1994){Tavani}, {Arons}, \& {Kaspi}}]{Tavani:1994qu}
{Tavani}, M., {Arons}, J., \& {Kaspi}, V.~M. 1994, \apjl, 433, L37

\end{thebibliography}

\appendix

\section{Doppler boosted inverse Compton emission on stellar photons}
 
 \begin{figure}
\centering
\resizebox{6.5cm}{!}{\includegraphics{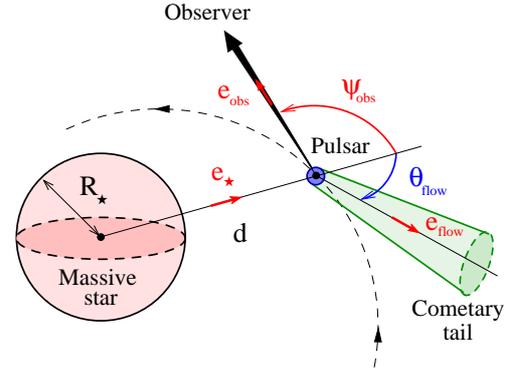}} 
\caption{Geometry of the binary + pulsar wind nebula flow system. The calculations assume that the massive star is point-like and that emission in the tail is limited to a small region at the pulsar location.}
\label{geometry}
\end{figure}

The star is approximated as a point source of photons and the electrons are confined in a very small region. The overall geometry and vectors are shown in Fig.~\ref{geometry}. In the point-like and mono-energetic approximation, the stellar photon density in the observer frame is
\begin{equation}
\frac{dn}{d\epsilon d\Omega}=n_0 \delta(\epsilon-\epsilon_0)\delta(\mu-\mu_0)
\end{equation}
where $\epsilon_0$ is the incoming photon energy and $\mu_0$ is the cosine of the angle between the incoming photon and the electron direction. Applying relativistic transforms to go to the comoving frame gives
\begin{equation}
\frac{dn^\prime}{d\epsilon^\prime d\Omega^\prime}=\Gamma^2 \left( 1-\beta \mathbf{e_\star}. \mathbf{e_{\rm flow}} \right)^2 \frac{dn}{d\epsilon d\Omega}={\cal D}_{\star}^{-2}\frac{dn}{d\epsilon d\Omega}
\end{equation}
Developing the Dirac functions leads to
\begin{equation}
\frac{dn^\prime}{d\epsilon^\prime d\Omega^\prime}=n^\prime_0 \delta(\epsilon^\prime-\epsilon^\prime_0)\delta(\mu^\prime-\mu^\prime_0)
\label{dn}
\end{equation}
with $n_0^\prime={\cal D}_{\star}^{-1} n_0$ and  $\epsilon^\prime_0={\cal D}^{-1}_{\star}\epsilon_0$. For inverse Compton scattering on an isotropic distribution of electrons in the comoving frame, $\mu^\prime_0\approx\mathbf{e^\prime_\star}.\mathbf{e^\prime_{\rm obs}}$ \citep{Dubus:2007oq}. The unit vector $\mathbf{e^\prime_\star}$ transforms in the comoving frame as
\begin{equation}
\mathbf{e}^\prime_\star=\frac{ \mathbf{e}_\star+\left[(\Gamma-1) (\mathbf{e}_\star . \mathbf{e_{\rm flow}})-\Gamma\beta\right]\mathbf{e_{\rm flow}} }{\Gamma ( 1 - \beta \mathbf{e}_\star . \mathbf{e_{\rm flow}})}
\end{equation}
The transform giving $\mathbf{e^\prime_{\rm obs}}$ is simply given by replacing $\mathbf{e}_\star$ with $\mathbf{e_{\rm obs}}$ above.  The dot product of the two vectors in the comoving frame simplifies to 
\begin{equation}
1-\mathbf{e^\prime_\star}.\mathbf{e^\prime_{\rm obs}}={\cal D}_{\rm obs} {\cal D}_{\star} (1-\mathbf{e_\star}.\mathbf{e_{\rm obs}})
\label{dot}
\end{equation}
The anisotropic inverse Compton scattering kernel in \citet{Dubus:2007oq} can then be used, with the photon density given in Eq.~\ref{dn} and with the direction given by $\mathbf{e^\prime_\star}.\mathbf{e^\prime_{\rm obs}}$. The resulting outgoing spectrum is then transformed back to the observer frame by using $\epsilon_1={\cal D}_{\rm obs} \epsilon_1^\prime$ and $F_\nu(\epsilon_1)={\cal D}^3_{\rm obs}  F^\prime_\nu(\epsilon^\prime_1)$ as discussed in \S3.1.  ${\cal D}_{\rm obs}$ is defined in Eq.~\ref{dobs} and $\epsilon_1$ is the outgoing photon energy.

For  inverse Compton emission by a power-law distribution of electrons in the Thomson regime, the spectrum in the comoving frame is given by 
\begin{equation}
F^\prime_\nu(\epsilon^\prime_1)=K n^\prime_0 (1-\mathbf{e^\prime_\star}.\mathbf{e^\prime_{\rm obs}})^{\frac{p+1}{2}}\left(\frac{\epsilon_1^\prime}{\epsilon_0^\prime}\right)^{\frac{1-p}{2}}
\label{eq1}
\end{equation}
where $p$ is the power-law index and $K$ is a constant. In this case, the spectrum seen by the observer is 
\begin{equation}
F_\nu(\epsilon_1)=K n_0  {\cal D}_{\star}^{-1-\alpha} {\cal D}_{\rm obs}^{3+\alpha}(1-\mathbf{e^\prime_\star}.\mathbf{e^\prime_{\rm obs}})^{\alpha+1}\left(\frac{\epsilon_1}{\epsilon_0}\right)^{-\alpha}
\end{equation}
so that, using the dot product in Eq.~\ref{dot},
\begin{equation}
F_\nu(\epsilon_1)=K n_0 {\cal D}_{\rm obs}^{4+2\alpha}(1-\mathbf{e_\star}.\mathbf{e_{\rm obs}})^{\alpha+1}\left(\frac{\epsilon_1}{\epsilon_0}\right)^{-\alpha}
\label{eq2}
\end{equation}
where $\alpha\equiv (p-1)/2$. This is identical to the expression found by \citet{1992A&A...256L..27D} and \citet{1993ApJ...416..458D} in the case of external scattering  by a jet propagating away from the seed photon source (an accretion disc). The formula in Eq.~\ref{eq1}-\ref{eq2} are formally only valid for Thomson scattering on an infinite power-law of electrons.

For completeness, the orbital separation $d$ is given by 
\begin{equation}
d=\frac{a(1-e^2)}{1+e\cos(\theta-\omega)}
\end{equation}
with the semi-major axis $a=(G M P^2_{\rm orb}/4\pi^2)^{1/3}$, $M$ the total mass, $e$ the eccentricity, $\theta$ the true anomaly and $\omega$ the periastron angle of the compact object.  If the flow is in the orbital plane where it makes an angle $\theta_{\rm flow}$ to the star - pulsar direction 
then
\begin{eqnarray}
& & \mathbf{e_{\rm obs}}.\mathbf{e_{\rm flow}} =-\sin(\theta+\theta_{\rm flow})\sin i \\
& & \mathbf{e_{\star}}.\mathbf{e_{\rm obs}}=-\sin\theta\sin i
\end{eqnarray}
 where $i$ is the inclination of the system.

  \end{document}